\documentclass[final,5p,times,twocolumn]{elsarticle}

\usepackage{lineno,hyperref}

\usepackage{float}

\modulolinenumbers[5]

\journal{Nucl. Instr. and Meth. Res. Sect. A}

\begin{document}

\begin{frontmatter}

\title{The PANDA DIRC Detectors}


\author[e]{E.~Etzelm\"{u}ller\corref{mycorrespondingauthor1}}
\ead{erik.etzelmueler@exp2.physik.uni-giessen.de}
\cortext[mycorrespondingauthor1]{Corresponding authors}
\author[a]{J.~Schwiening\corref{mycorrespondingauthor1}}
\ead{j.schwiening@gsi.de}
\author[a,b]{A.~Ali}
\author[a]{A.~Belias}
\author[a]{R.~Dzhygadlo}
\author[a]{A.~Gerhardt}
\author[a,2]{M.~Krebs}
\author[a]{D.~Lehmann}
\author[a,b]{K.~Peters}
\author[a]{G.~Schepers}
\author[a]{C.~Schwarz}
\author[a]{M.~Traxler}
\author[c]{L.~Schmitt}
\author[d]{M.~B\"{o}hm}
\author[d]{A.~Lehmann}
\author[d]{M.~Pfaffinger}
\author[d]{S.~Stelter}
\author[d]{F.~Uhlig}
\author[e]{M.~D\"{u}ren}
\author[e]{K.~F\"{o}hl}
\author[e]{A.~Hayrapetyan}
\author[e]{K.~Kreutzfeld}
\author[e]{J.~Rieke}
\author[e]{M.~Schmidt}
\author[e]{T.~Wasem}
\author[f]{C.~Sfienti}

\address[a]{GSI Helmholtzzentrum f\"ur Schwerionenforschung GmbH, Darmstadt, Germany}
\address[b]{Goethe University, Frankfurt a.M., Germany}
\address[c]{FAIR, Facility for Antiproton and Ion Research in Europe, Darmstadt, Germany}
\address[d]{Friedrich Alexander-University of Erlangen-Nuremberg, Erlangen, Germany}
\address[e]{II. Physikalisches Institut, Justus Liebig-University of Giessen, Giessen, Germany}
\address[f]{Institut f\"{u}r Kernphysik, Johannes Gutenberg-University of Mainz, Mainz, Germany}

\begin{abstract}
The PANDA experiment at the future Facility for Antiproton and Ion Reasearch (FAIR) will address 
fundamental questions of hadron physics with unprecedented precision. 
To reach this goal excellent Particle Identification (PID) is essential over a large range of particle momenta and solid angles. 
Most of the phase space will be covered by two innovative DIRC (Detection of Internally Reflected Cherenkov light) detectors. 
The Endcap Disc DIRC and Barrel DIRC will cover the polar angle range from 5 to 22$^\circ$ and 22 to 140$^\circ$, respectively. 
Both detectors rely on high precision optical components, lifetime-enhanced Microchannel Plate PMTs (MCP-PMTs), and fast readout electronics.
\end{abstract}


\end{frontmatter}

\begin{figure*}[t]
  \centering
    \includegraphics[width=0.8\textwidth]{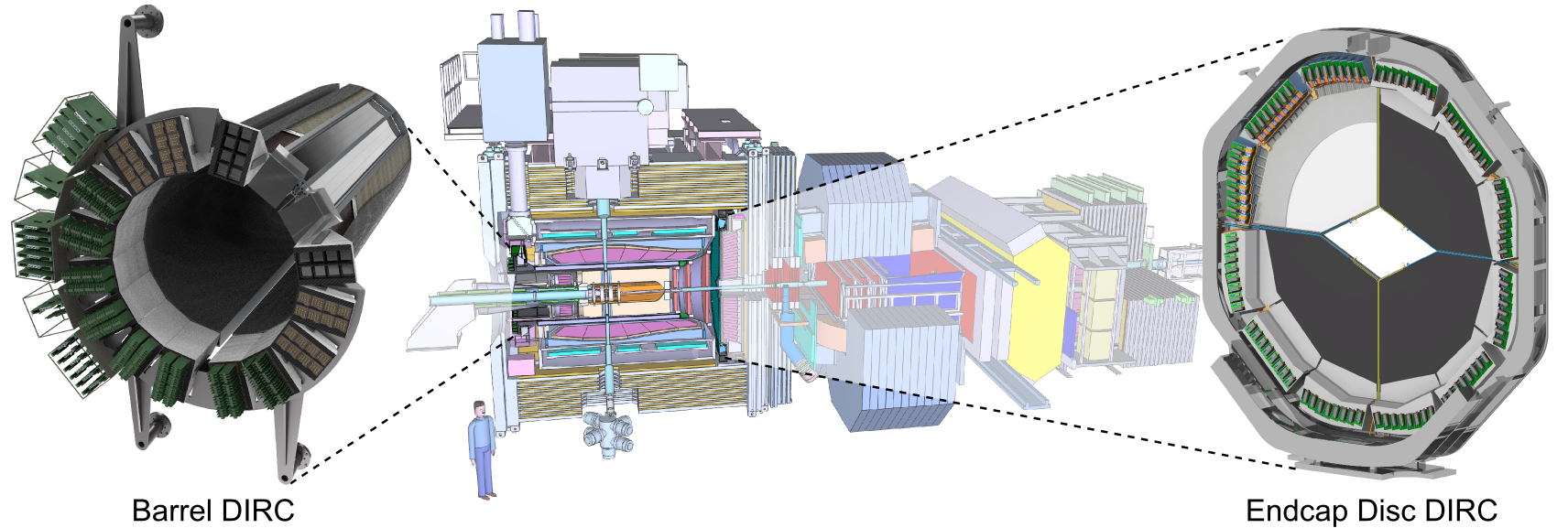}
    \caption{Side view of the PANDA experiment, highlighting the locations of the two DIRC detectors.}
\label{fig:panda_dircs} 
\end{figure*}


\section{PANDA at FAIR}

The PANDA (AntiProton Annihilation at Darmstadt) experiment will be one of the four pillars of the future Facility for Antiproton and Ion Research (FAIR) which is currently under construction at Gesellschaft f\"ur Schwerionenforschung (GSI) in Darmstadt, Germany. Antiprotons with momenta ranging from 1.5 to 15\,GeV/c will collide with a fixed proton or hypernuclear target to study various topics involving weak and strong forces, exotic states, and the structure of hadrons. The PANDA detector will provide precise tracking, energy measurement, and efficient particle identification (PID). PID of charged hadrons in the target spectrometer of the PANDA experiment will be realized primarily by two DIRC counters: the Endcap Disc DIRC for polar angles between 5 and 22$^\circ$ and the Barrel DIRC for angles between 22 and 140$^\circ$ (see Fig.~\ref{fig:panda_dircs}). 

\section{DIRC principle}

In contrast to conventional Ring Imaging Cherenkov (RICH) detectors, which use gas or aerogel to produce the Cherenkov light, DIRC detectors are based on solid radiator materials, in the form of bars or plates, with a much higher refractive index. This allows to operate the RICH technology in a region of lower particle momenta and enables the construction of thin detectors with a depth of only few centimeters along the particle trajectory. The key technology of each DIRC is the fabrication of highly parallel and precisely polished optical surfaces which conserve the Cherenkov angle information during total internal reflection, which traps part of the produced Cherenkov photons and guides them towards the readout region. For this purpose synthetic fused silica is typically used for the radiator due to its radiation hardness and machinability. The detection of the Cherenkov light is based on single photon sensors which, in the case of PANDA, must provide a high lifetime, radiation hardness and have to work inside a magnetic field of up to 1\,T in the readout region.


\section{The PANDA Endcap Disc DIRC}

\subsection{Detector Design}

\begin{figure} [hbt]
    \centering
    \includegraphics[width=0.85\columnwidth]{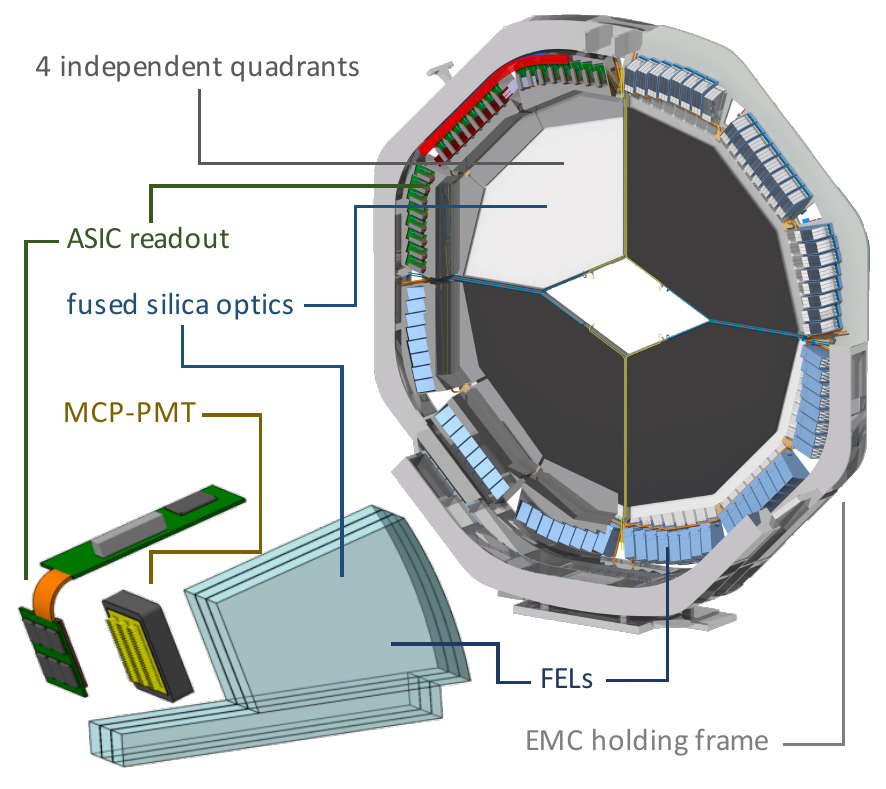}
    \caption{CAD image of the complete Endcap Disc DIRC and the inside of a Readout Module (ROM), indicating key components.}
    \label{fig:edd_cad}
\end{figure}

The Endcap Disc DIRC (EDD) will provide particle identification of charged pions from kaons for momenta up to 4\,GeV/c with a separation power of at least 3 standard deviations. Covering the forward endcap region of the PANDA experiment, the requirements for the detector differ from other designs to date. The detector is divided into four independent quadrants which together form a dodecagon with a central rhombic opening for the acceptance of the PANDA Forward Spectrometer. Each quadrant consists of a radiator and 24 Readout Modules (ROMs). Each ROM contains three bars, three Focusing Elements (FELs), a Microchannel Plate Photomultiplier Tube (MCP-PMT), and a Frontend Electronics board (see also Fig.~\ref{fig:edd_cad}). 

\begin{figure} [h]
  \centering
    \includegraphics[width=0.85\columnwidth]{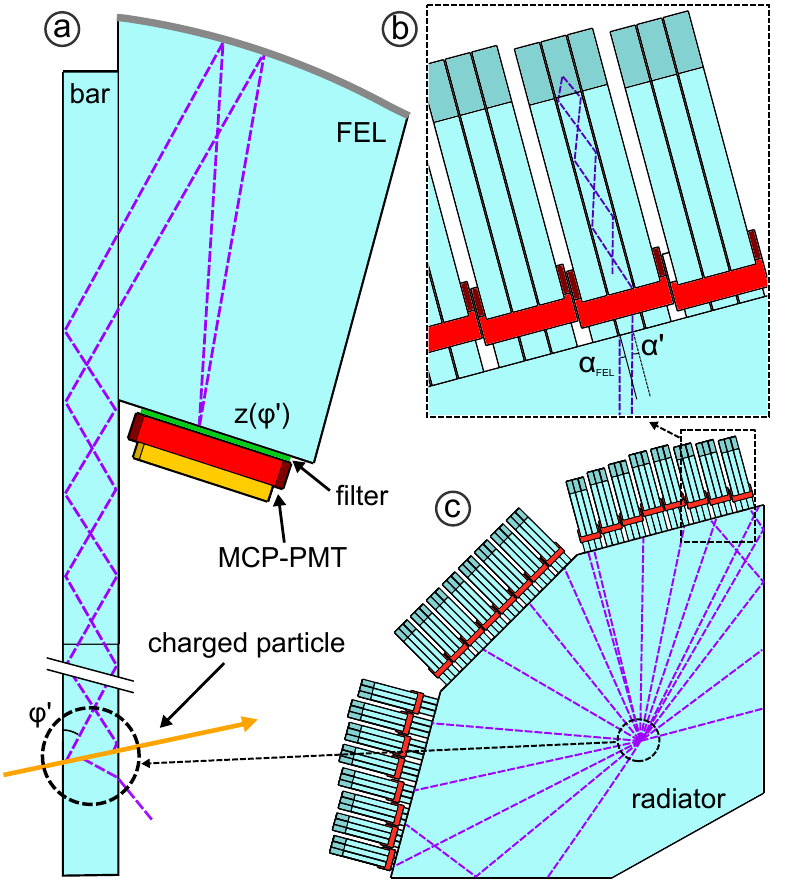}
    \caption{(a) A charged particle traverses the radiator and emits Cherenkov photons (dashed lines). Photons which fulfill the condition for total internal reflection propagate towards the ROMs where they enter a bar and subsequently a Focusing Element (FEL) where they are focused on a MCP-PMT. (b) The azimuthal angle is determined from the position of the bar where the photon enters the ROMs. (c) shows different light paths inside the radiator.}
\label{fig:edd_wp}
\end{figure}

Radiator, bars, and FELs form the optical system of the detector and are made of synthetic fused silica. 
The thickness of a radiator amounts to 20\,mm at a maximum diameter of 1.5\,m. To minimize photon loss and conserve the angle information the surface roughness is below 1\,nm RMS with a Total Thickness Variation (TTV) below 15\,$\mu$m. For manufacturing reasons, and due to space limitations, the bars extend the radiator volume. They will be glued to the outer rim of the radiator plate in order to avoid the need for optical in situ alignment and to reduce the needed mechanical support structure. To improve the focusing quality and reduce photon loss the coupling of the bars and FELs is done via optical contact bonding. The FELs transform the angle information into a position information by internally focusing the Cherenkov photons via an aluminum coated cylindrical surface (see Fig.~\ref{fig:edd_wp}).

The shape of the FELs is optimized to match the resolution of the MCP-PMTs, which have a pitch size of 0.5\,mm or smaller, and at the same time align them with respect to the magnetic field of roughly 0.8\,T to provide sufficient gain for single photon detection. The MCP-PMTs are 2-inch tubes with a customized anode containing either $3\times 100$ or $6\times 128$ pixels. Measurements inside a magnetic field have shown that charge sharing between adjacent pixels will only occur at the edges of the pixels as the (secondary) photoelectrons are guided by the magnetic field lines \cite{julian_phd}.\\
In order to extend the life time of the MCP-PMTs, which is proportional to the accumulated charge on the photocathode, which can reach 7\,C/cm$^2$ or more during the lifetime for the EDD, and to mitigate the contribution of the chromatic error, a long pass wavelength filter will be applied. Simulations have shown that an optimal cut-off wavelength is at around 350\,nm which can be realized either by an optical filter in front of the MCP-PMT or a modified photocathode~\cite{michael_rich16}.\\
The analog single photon signal originating from the MCP-PMTs will be readout and digitized by the TOFPET2 ASIC by PETsys with 64 channels per ASIC \cite{petsys}. The ASICs will be bonded to a flex PCB which forms the Frontend Electronics Board together with an FPGA and a versatile link option. In total the system will be able to readout a maximum of 36,864 channels.

\subsection{Reconstruction and Performance}

The reconstruction of the Cherenkov angle is based on a three dimensional measurement: the pixel position on the MCP-PMT ($z(\phi^\prime)$), the position of the FEL and the time of propagation. The latter is not used directly in the reconstruction algorithm but resolves ambiguities of overlapping patterns. The resolution of $z(\phi^\prime)$ is limited by the pixel size and the resolution of the FELs as well as the width of the bars as indicated in Fig.~\ref{fig:edd_wp}\,(b) as the exact azimuthal position of the Cherenkov light entering the bar is unknown.\\
Taking into account the momentum vector of the particle emitting the Cherenkov photons, the Cherenkov angle can be calculated analytically for each pixel hit. On average 22 photons are detected for each charged particle track. For geometrical reasons the performance depends on the particle track position and 
ranges between 1.1 and 2.2\,mrad at 4\,GeV/c which translates into a separation power between 3 and 6 standard deviations~\cite{mustafa_jinst}. 

\begin{figure} [h]
  \centering
    \includegraphics[width=0.9\columnwidth]{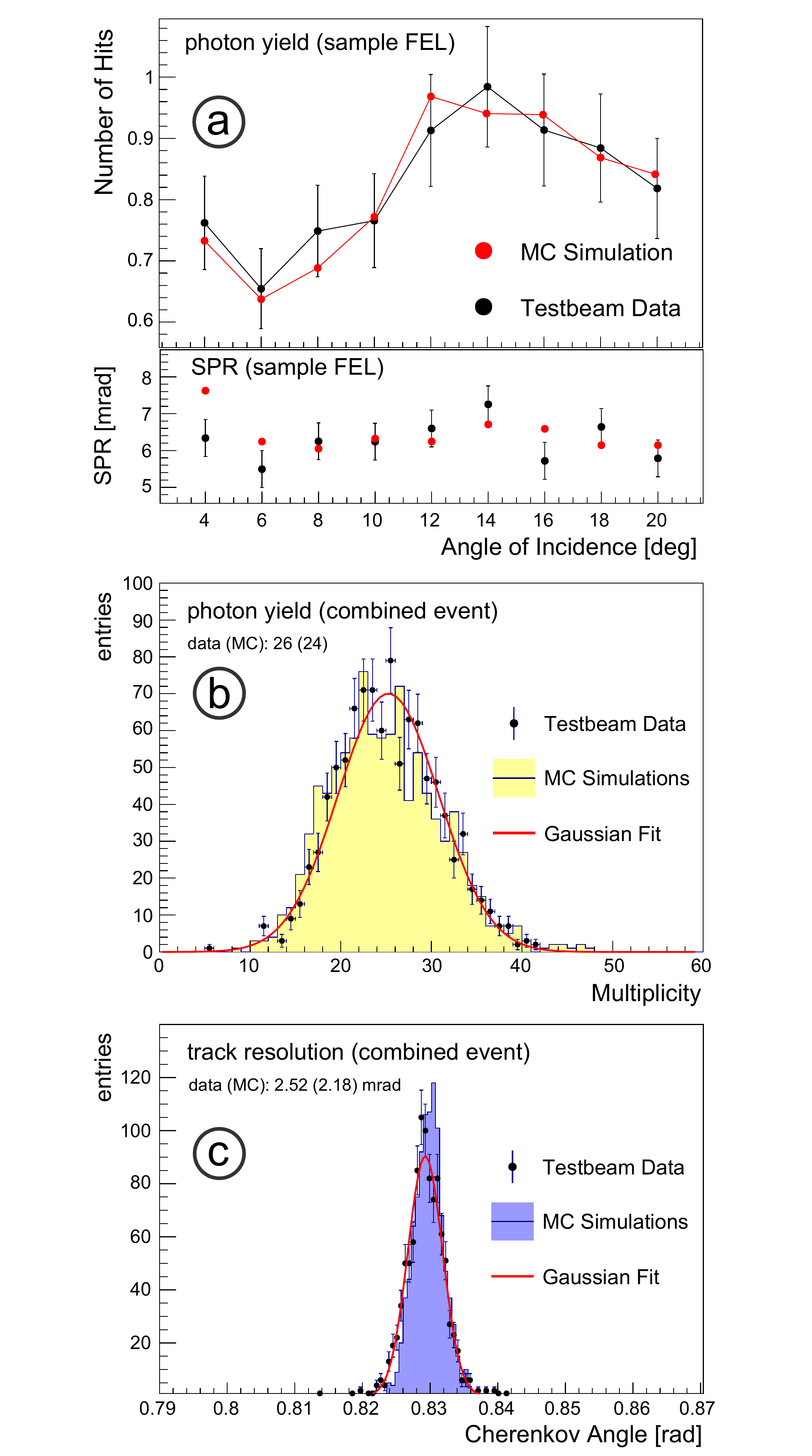}
    \caption{(a) photon yield and single photon resolution (SPR) for a single FEL for an electron beam at DESY during the 2016 test beam campaign. (b) and (c) show the photon yield and track resolution for the combined event analysis, respectively.}
\label{fig:edd_results}
\end{figure}

\subsection{Prototype Results}

In the course of the development of the detector, individual prototypes of all previously described detector components have been produced and tested including custom made setups for quality assurance and assembly \cite{julian_phd,erik_phd}.\\
Since 2015 all detector prototypes, which were tested in hadron and electron beams at CERN and DESY, consisted of a $500\times 500\times 20\,\mathrm{mm}^3$ radiator, focusing optics made of synthetic fused silica, and high-resolution MCP-PMTs. The readout electronics and mechanics were improved for each test beam campaign, featuring a TOFPET2 ASIC readout and 3D-printed ROM mechanics in the latest version.\\
Because of the reduced size of the radiator and the limited number of available ROMs for the EDD prototype, the analysis of the test beam data focused on the single photon resolution (SPR) and photon yield, comparing them to Monte Carlo predictions (see Fig.~\ref{fig:edd_results}\,(a)). The Monte Carlo data can subsequently be extrapolated from the test beam geometry to the final geometry. A SPR value of 6.06\,mrad, measured using a 3\,GeV/c electron beam at DESY in 2016, corresponds to a SPR for pions in the full EDD detector of 1.8\,mrad at the same momentum~\cite{mustafa_jinst}.\\
Another approach was taken for the so called combined event analysis, where the data taken during a vertical scan for a single FEL was combined to simulate a prototype equipped with 30 FELs (or 10 ROMs). For each combined event the photon yield and the average so-called track resolution was calculated and compared to Monte Carlo predictions. Fig.~\ref{fig:edd_results}\,(b) and (c) show the good agreement between the test beam data and the simulation with a measured combined track resolution of 2.52\,mrad (2.18\,mrad for the Monte-Carlo simulation)~\cite{klaus_jinst}. 


\section{The PANDA Barrel DIRC}

\subsection{Detector Design}

\begin{figure} [h]
	\centering
	\includegraphics[width=0.99\columnwidth]{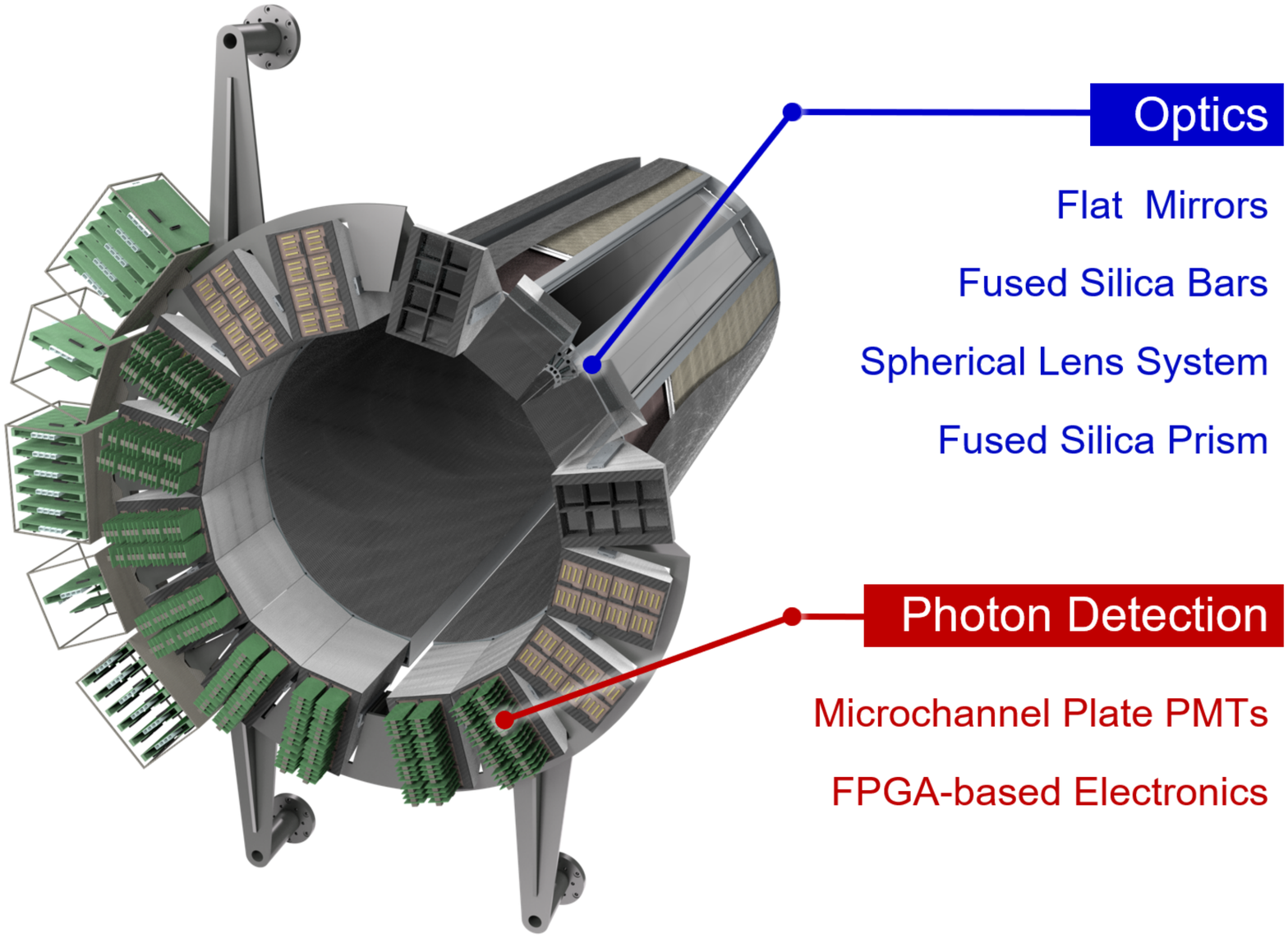}
	\caption{Schematic of the PANDA Barrel DIRC.
	}
	\label{fig:barrel-1}
\end{figure}

The Barrel DIRC is designed to provide clean separation of charged pions from kaons with 3 
standard deviations or more for momenta up to 3.5\,GeV/c. 
The concept, shown in Fig.~\ref{fig:barrel-1}, is based on the successful BaBar DIRC 
detector~\cite{adam2005} as well as key results from the R\&D for the SuperB 
FDIRC~\cite{superb:dirc1}, with several important improvements, such as fast photon timing 
and a compact imaging region.

The design is described in detail in the Technical Design Report~\cite{Barrel-TDR-arXiv}. 
Sixteen optically isolated sectors, each comprising a bar box and a synthetic fused silica prism, 
surround the beam line in a 16-sided polygonal barrel with a radius of 476\,mm. 
Each bar box contains three synthetic fused silica bars of 17\,mm thickness, 53\,mm width, 
and 2400\,mm length, placed side-by-side, separated by a small air gap. 
A flat mirror is attached to the forward end of each bar to reflect photons towards the readout end, 
where they are focused by a 3-layer spherical lens on the back of a $300$~mm-deep solid prism, 
made of synthetic fused silica, serving as expansion volume (EV).
An array of 2~$\times$~4 lifetime-enhanced MCP-PMTs~\cite{lehmann2016}, each with 
8~$\times$~8 pixels of about 6.5~$\times$~6.5\,mm$^2$ size, is placed at the back 
surface of the prisms to detect the photons and measure their arrival time on a total of 
8192 pixels with a precision of about $100$\,ps in the magnetic field of approximately 1\,T.

The readout of the MCP-PMTs is based on the TRB-technology~\cite{trb3}, combined with the  
FPGA-based DiRICH front-end amplification and discrimination module~\cite{dirich}. 
The MCP-PMTs are mounted on a highly-integrated backplane, which provides the high voltage 
to the sensors as well as low voltages and signal routing to the DiRICH cards. 
By measuring the photon arrival time as well as the Time-over-Threshold of the signal, the 
performance of the sensors can be monitored and time-walk effects can be corrected, 
improving the precision of the single photon timing.

All major mechanical components will be be built from aluminum alloy and 
Carbon--Fiber--Reinforced Polymer (CFRP) to minimize the material budget and weight 
and to maximize the stiffness.

The PANDA Barrel DIRC will be the first DIRC counter utilizing lens focusing. 
The innovative design of the optics produces a flat image, matching the shape of the 
back surface of the fused silica prism. 
This is achieved by a combination of focusing and defocusing elements in a spherical triplet 
lens made from one layer of lanthanum crown glass 
between two layers of synthetic fused silica. 
This 3-layer lens works without any air gaps, minimizing the photon loss that would 
otherwise occur at the transition from the lens to the expansion volume.

\subsection{Reconstruction and Performance}

The performance of the PANDA Barrel DIRC design was evaluated using a detailed physical simulation, implemented in Geant4~\cite{Geant4}, and validated using complex prototypes in particle beams at GSI and CERN. The Geant simulation is tuned to wavelength-dependent properties of the materials used and the most important Cherenkov photon loss processes: the photon transport efficiency during total internal reflections~\cite{babar:dirc1} and the quantum and collection efficiency of MCP-PMTs~\cite{fred02}.

Two reconstruction approaches were developed to evaluate the detector 
performance~\cite{RICH14_sim}. 
The geometrical reconstruction is based on the approach used by the BaBar DIRC~\cite{adam2005}. 
The value of the Cherenkov angle of each detected photon is reconstructed using the relative position 
of the bar and the MCP-PMT pixel, in combination with the particle direction, and PID is performed in a track-by-track maximum likelihood test. 
The algorithm can also determine the Cherenkov angle per particle and the photon yield, as well 
as the single photon Cherenkov angle resolution, important figures of merit in evaluating the performance 
of a prototype design in test beams and comparing it to other Cherenkov detectors. 
While the geometrical method relies primarily on the position of the detected photons, the time-based imaging approach, based on the method used by the Belle II TOP~\cite{staric}, utilizes both the position 
and time information with comparable precision, and directly performs the maximum likelihood test 
based on the photon arrival time in each pixel.

\subsection{Prototype Results}

The performance of a number of PANDA Barrel DIRC system prototypes, featuring different radiator 
and prism geometries, focusing options, and readout electronics, was determined using particle 
beams at GSI and CERN from 2011--2017 to validate the design and the simulation results.

\begin{figure} [h]
	\centering
	\includegraphics[width=0.85\columnwidth]{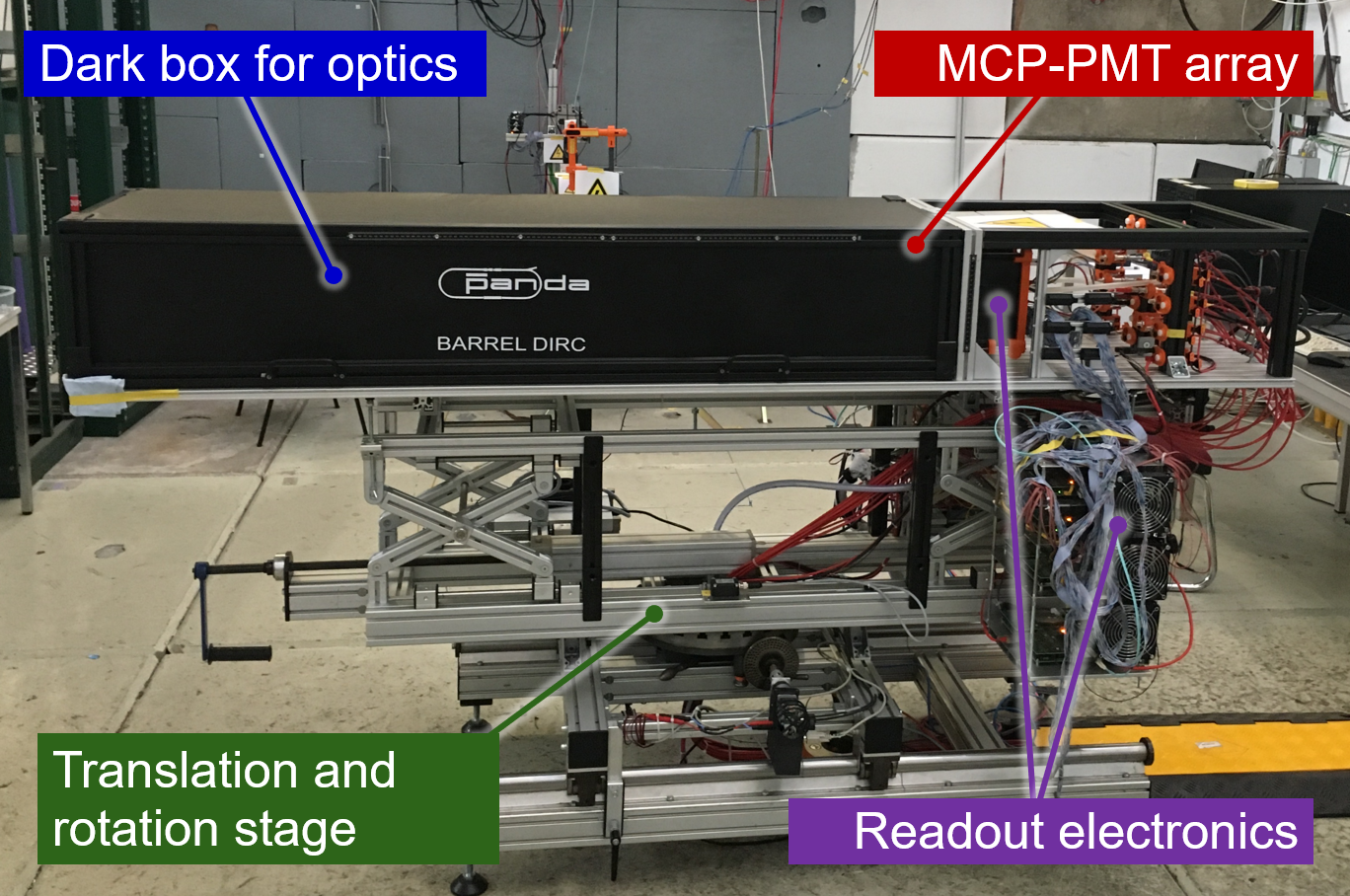}
	\caption{Photograph of the 2017 prototype in the CERN T9 beam line.
	}
	\label{fig:barrel-2}
\end{figure}

The prototype tested at the T9 beamline at the CERN PS in 2017, shown in Fig.~\ref{fig:barrel-2}, 
featured all critical elements of the final PANDA Barrel DIRC design: 
A narrow fused silica radiator bar (17.1 $\times$ 35.9 $\times$ 1200.0\,mm$^3$) coupled on 
one end to a flat mirror, on the other end to a 3-layer spherical focusing lens, a fused silica prism 
as expansion volume (with a depth of 300\,mm and a top angle of 38$^\circ{}$), 
an array of 2$\times$4 MCP-PMTs, and 600 readout electronics channels using TRBs in 
combination with FPGA-based amplification and discrimination cards 
(PADIWA)~\cite{{cardinali:padiwa}}, mounted directly on the MCP-PMTs.

The time difference measured by two time-of-flight stations was used to cleanly tag an event 
as pion or as proton. 
The experimental data showed good agreement of the Cherenkov hit patterns with simulation, 
both in the pixel space and in the photon hit time space.

\begin{figure}[htb]
	\centering
	{\includegraphics[width=0.9\columnwidth]{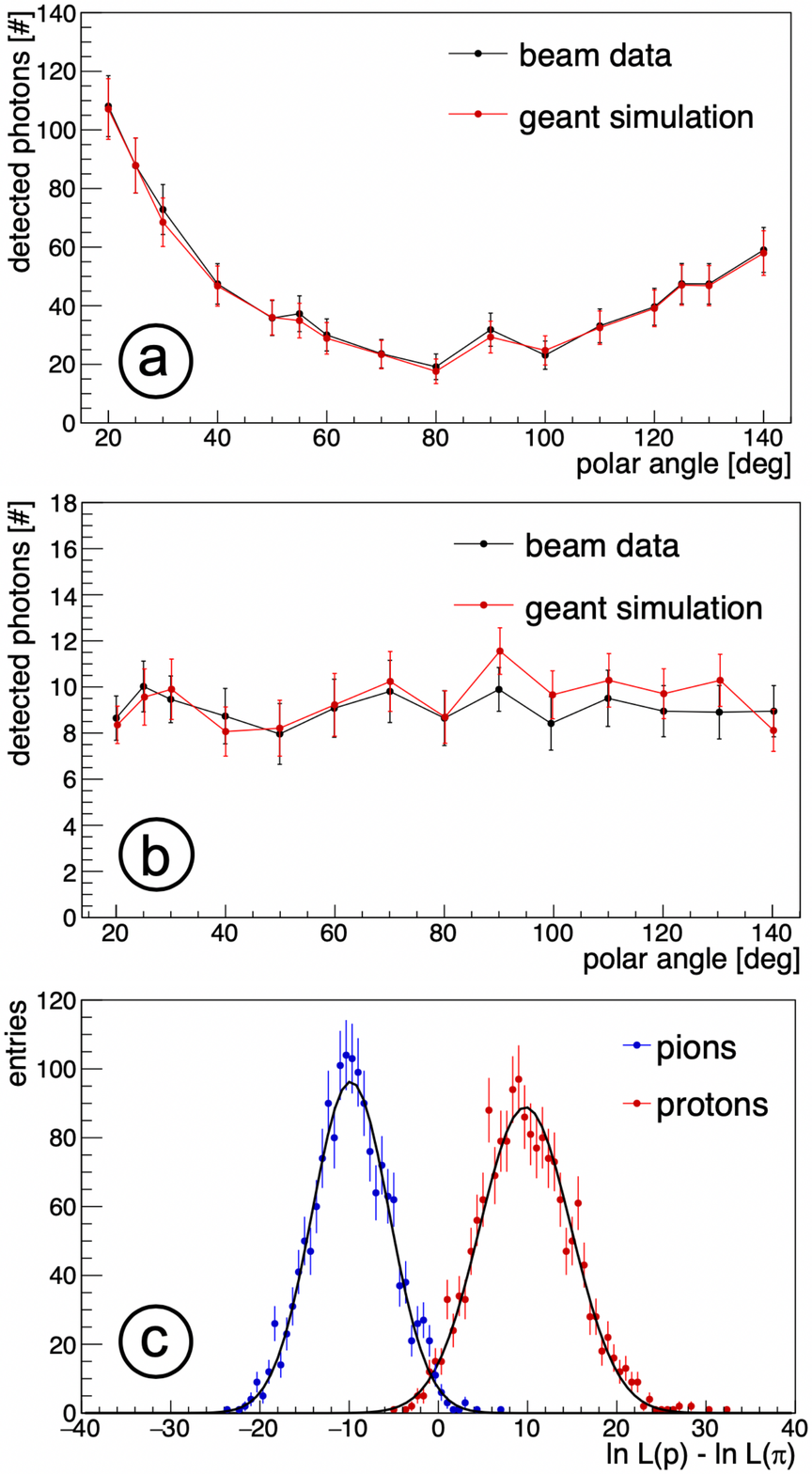}}
	\caption{Performance of the PANDA Barrel DIRC prototype at the CERN PS in 2017 
		for a beam momentum of 7\,GeV/$c$.\newline
		Photon yield (a) and single photon Cherenkov angle resolution (SPR) (b) from the geometric reconstruction as a function of the track polar angle for tagged protons. The error bars correspond to the RMS of the distribution in each bin. \newline
		Proton-pion log-likelihood difference distributions (c) for proton-tagged (red) and pion-tagged (blue) beam events from time-based imaging reconstruction for a polar angle of $20^{\circ}$.  
	}
	\label{fig:barrel-3}
\end{figure}

Examples of the performance of the PANDA Barrel DIRC prototype with a narrow bar and a 
3-layer spherical lens at the CERN PS in 2017 are shown in Fig.~\ref{fig:barrel-3}. 
For tagged protons at 7\,GeV/$c$ momentum the number of Cherenkov photons ranges from 
20 to 110 and the single photon Cherenkov angle resolution (SPR) varies between 8--11\,mrad, 
both in good agreement with simulation. 
The log-likelihood difference distributions obtained using time-based imaging reconstruction 
for tagged pions and protons with a polar angle of $20^{\circ}$ and a momentum of  7\,GeV/$c$ 
is shown in Fig.~\ref{fig:barrel-3}\,(c). 
The separation power determined from the Gaussian fits is 4.2$\pm$0.1\,s.d., corresponding 
to a $\pi/K$ separation power of 4.4$\pm$0.1\,s.d. at a momentum of 3.5\,GeV/c. 
Simulation can be used to extrapolate this value to the performance of the fully equipped 
PANDA Barrel DIRC. 
Assuming that the expected technical specifications of the MCP-PMTs, lenses, and readout 
electronics, in particular the photon detection efficiency and timing precision, are achieved as planned,
the result extrapolates to a $\pi/K$ separation power of about 7.7\,s.d. at $25^{\circ}$ and 
3.5\,GeV/$c$ in PANDA, exceeding the PANDA PID requirements.


\section{Summary and Outlook}

The PANDA Endcap Disc DIRC and the Barrel DIRC have shown significant progress over the past few years. The technical designs were completed and the performance of prototypes validated with complex prototypes in particle beams at CERN and DESY. Both DIRC detectors are now transitioning to the construction phase with component fabrication starting in early 2019 and installation into PANDA scheduled for 2023/2024,followed by commissioning with protons soon after and first physics with antiprotons in 2025.

\section*{Acknowledgements}

This work was supported by the Bundesministerium f\"ur Bildung und Forschung (BMBF), HGS-HIRe, HIC for FAIR, and by the EIC detector R\&D (eRD14) fund, managed by Brookhaven National Lab. We thank GSI, DESY and CERN staff for the opportunity to use the beam facilities and for their on-site support as well as the RICH2018 organizers in Moscow for an outstanding meeting.


\end{document}